# Dislocation-enhanced piezoelectric catalysis of KNbO₃ crystal for water splitting

Hanyu Gong[1a], Jiawen Zhang[2a], Yan Zhao[1], Shan Xiang[1], Xiang Zhou[1], Oliver Preuß[3], Wenjun Lu[2*], Yan Zhang[1*], Xufei Fang[4*]

[1]State Key Laboratory of Powder Metallurgy, Central South University, Changsha, 410083, China

[2]Department of Mechanical and Energy Engineering, Southern University of Science and Technology, Shenzhen 518055, China

[3]Department of Materials and Earth Sciences, Technical University of Darmstadt, 64287 Darmstadt, Germany

[4]Institute for Applied Materials, Karlsruhe Institute of Technology, 76131 Karlsruhe, Germany

[a]These two authors contributed equally to this work.

*Correspondence: luwj@sustech.edu.cn (W. L.); yanzhangcsu@csu.edu.cn (Y. Z.); xufei.fang@kit.edu (X. F.)

**Abstract**

Dislocations in oxides with ionic/covalent bonding hold potential of harnessing versatile functionalities. Here, high-density dislocations in a large plastic zone in potassium niobate (KNbO₃) crystals are mechanically introduced by room-temperature cyclic scratching to enhance piezocatalytic hydrogen production. Unlike conventional energy-intensive, time-consuming deformation at high temperature, this approach merits efficient dislocation engineering. These dislocations induce local strain and modify the electronic environment, thereby improving surface reactivity and charge separation, which are critical for piezocatalysis. This proof-of-concept offers a practical and sustainable alternative for functionalizing piezoelectric ceramics. Our findings demonstrate that surface-engineered dislocations can effectively improve the piezocatalysis, paving the way for efficient and scalable piezocatalytic applications.

*Graphical Abstract*

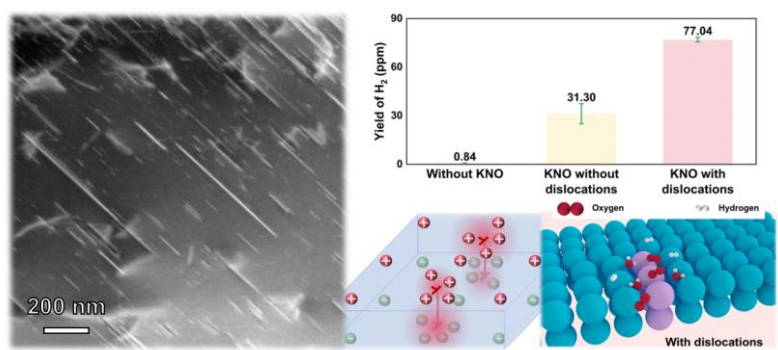

Dislocations in ceramics have been recently discovered as a renewed tool for tuning the mechanical properties such as plasticity[1], strength[2], and fracture resistance[3]. However, their role in functional properties, especially in catalysis, has fallen short for exploration. Dislocations, atomic line distortion in crystalline solids, can not only introduce local high strain field[4] but also alter the electronic environment within ceramics with ionic/covalent bonding. These features have the potential to influence catalytic performance by improving surface reactivity and facilitating charge separation[5]. These effects are especially valuable in catalytic reactions, where surface-active sites are crucial for the adsorption and transformation of reactants.

Piezocatalysis[6], which utilizes the piezoelectric effect to drive catalytic reactions under mechanical stress, offers several advantages over traditional catalytic methods such as photocatalysis, thermo-catalysis, and electrocatalysis. Compared to photocatalysis[7], which is limited by light availability and wavelength constraints, piezocatalysis can operate efficiently in the absence of light and is not restricted to specific wavelengths. Thermo-catalysis[8] requires high temperatures and substantial energy input. Piezocatalysis works at room temperature, significantly reducing energy consumption. Furthermore, it does not rely on external electrical sources as electrocatalysis does, simplifying the system and reducing overall energy demand[9]. These factors make piezocatalysis an attractive and energy-efficient alternative, particularly in applications like water splitting and hydrogen production[10]. Our previous research[11] on barium titanate ($BaTiO_3$) single crystals has shown that introducing dislocations can enhance piezocatalytic hydrogen production. However, this typical method for introducing dislocations into $BaTiO_3$ involves high-temperature mechanical imprinting[12], which is energy-intensive and results in dislocations being introduced deep within the crystal structure. Most of these internal dislocations are unlikely to effectively participate in surface catalytic reactions, which are essential for piezocatalysis. Furthermore, the high-temperature imprinting process requires expensive bulk crystals, which may also form cracks during compression and cooling, reducing the structural integrity and stability of catalysts. Despite these limitations, the potential of dislocations to improve piezocatalytic performance remains significant, and thus, more efficient and cost-effective methods are needed to introduce dislocations, preferably at room temperature.

In this study, we explore an approach for enhancing the piezocatalytic performance of potassium niobate ($KNbO_3$) single crystals by introducing dislocations via mechanical scratching at room temperature. Unlike $BaTiO_3$, which requires high-temperature processing for dislocation introduction, $KNbO_3$ is capable of room-temperature deformation[13], allowing for the controlled introduction of dislocations at the material surface without the risk of fracture[2]. This method offers an energy-efficient alternative to traditional techniques and enhances the material's surface reactivity, leading to improved piezocatalytic hydrogen production. The findings demonstrate that surface-localized dislocations in KNbO$_3$ effectively improve its catalytic activity, offering a promising approach for piezocatalysis without the need for the costly high-temperature treatment. This work aims to provide a more sustainable and efficient strategy for functionalizing piezoelectric ceramics in catalytic applications.

Undoped KNbO$_3$ (KNO) single crystals (FEE GmbH, Idar-Oberstein, Germany), grown by top-seeded solution method, were used for this study. At room temperature, KNO has an orthorhombic crystal structure. As temperature increases it undergoes two phase transitions from orthorhombic to tetragonal at 225 °C, then to cubic at 435 °C[14]. The crystallographic directions used later are the pseudo-cubic directions[14,15] for consistency of room-temperature slip system definition as in other perovskite oxides such as SrTiO$_3$ and KTaO$_3$, both of which have cubic structure at room temperature. The samples were cut into the dimension of about 5 mm × 5 mm × 1 mm with the large surface being (001). The (001) surfaces were sequentially ground with wet grinding papers (P800, P1200, P2500, and P4000, QATM, Mammelzen, Germany). Then the samples were further polished (Phoenix 4000, Buehler, Lake Bluff, IL, USA) with diamond polishing paste (particle sizes being 6, 3, 1, and 1/4 µm) for 30 min each. A final step of vibrational polishing was adopted using OPS polishing solution containing ~ 50 nm colloidal silica particles for 20 h. This detailed procedure has proven to be effective in removing grinding/polishing induced unintended near-surface damage.

To intentionally engineer high-density dislocations, the cyclic Brinell indenter scratching method[16] was adopted at room temperature using a universal hardness testing machine (Finotest, Karl-Frank GmbH, Weinheim-Birkenau, Germany), which is mounted with a spherical indenter made of hardened steel, with a diameter of 2.5 mm. The Brinell indenter was brought into contact with the sample (100) surface with a load of 0.8 kg to slide along the <110> direction at a velocity of 0.5 mm s$^{-1}$, controlled by a single-axis piezo stage (PI Instruments, Karlsruhe, Germany). To increase the dislocation density inside the scratch tracks, 10 passes (10×) were used for each scratch track[17]. Silicone oil was used as a lubricant during the scratch tests. The region affected by the plastic deformation (single scratch track) has a width of ~150 µm and a depth of tens of micrometers. To enlarge the plastic zone, multiple parallel scratch tracks were overlapped on the (001) surface to create a region of about 3 mm × 3 mm dislocation zone (see Supplementary Materials **Figure. S1**).

To characterize the generated dislocations via Brinell indenter scratching, we used electron channeling contrast imaging (ECCI) within an SEM (scanning electron microscopy, Tescan MIRA3-XM, Brno, Czech Republic) equipped with a four-quadrant solid-state BSE detector (DEBEN, Woolpit, UK). The acceleration voltage was 15 kV with a working distance of 8 mm. A carbon layer of approximately 10 nm thickness was sputtered on top of the sample to reduce the surface charging effect. As ECCI is a surface imaging technique, for in-depth and cross-sectional visualization of the dislocations, TEM (transmission electron microscope) specimens were prepared inside the plastic zone, along the <100> orientation, using a dual-beam focused ion beam (FIB, Helios Nanolab 600i, FEI, Hillsboro, USA). Annular dark field scanning TEM (ADF-STEM) imaging was performed on a TEM instrument (FEI Talos F200X G2, Thermo Fisher Scientific, USA) at an operating voltage of 200 kV. A probe semi-convergence angle of 10.5 mrad and inner and outer semi-collection angles of 23-55 mrad were used in the ADF-STEM imaging. Atomic scale TEM analysis was performed on a double aberration-corrected TEM (TitanThemis G2, FEI, USA) operating at 300 kV. A probe semi-convergence angle of 17 mrad and an inner and outer semi-collection angle of 38–200 mrad were used for high angle annular dark field STEM (HAADF-STEM) imaging.

Piezoelectric hydrogen production was conducted in a closed glass vessel with the volume of 713 ml. $KNbO_3$ single crystal was placed into a glass vessel containing 35 ml pure water, without any sacrificial agents. After absorption in water for 30 min, nitrogen ($N_2$) was passed through the solution with the gas flow rate of 8 L/h for 15 min to obtain an inert atmosphere. Then the instantaneous yield of $H_2$ ($Y_H$) at intervals of 30 min was measured by the gas chromatograph (UATEC-6600, Fanwei (Shanghai) Analytical Instruments Co., LTD) equipped with thermal conductivity detector (EHP15887, Valco Instruments Co. Inc, USA). The piezocatalytic $H_2$ production rate was calculated by Eq. (1)[18]:

$$v = Y_H V / V_m t \qquad (1)$$

where $V$ is the total gas volume of the closed glass vessel, and $V_m$ is the ideal molar volume of gas which is 22.4 L/mol, and $t$ is reaction time.

ADF-STEM analysis reveals the dislocation distribution and dislocation structure of the (001) $KNbO_3$ single crystal with 10× scratching. As displayed in **Figure 1a**, the dislocation density reaches to ~$10^{14}$ m$^{-2}$ after 10× scratching, and the dislocations are mainly composed of long 45° segments (marked by the blue arrows in **Figure 1b**) and irregular segments (marked by the green arrows in **Figure 1b**). Two-beam analysis along different diffraction conditions was then employed to confirmed the Burgers vector of the dislocation lines generated by cyclic scratching. As demonstrated in **Figure 1d**, the 45° segments were invisible along the diffraction vector $\boldsymbol{g} = 1\bar{1}0$, while visible along the other diffraction conditions in **Figure 1**. Hence, the dislocation Burgers vector $\boldsymbol{b}$ yields [110]. The dislocation line vector $\boldsymbol{t}$ was determined to [110] by the dislocation analysis along the <001> zone axis. Since the dislocation Burgers vector is parallel to the line vector, these 45° dislocation segment are confirmed as screw-type dislocations, and belong to the <110>{110} slip system[19]. The irregular dislocation segments were invisible along the diffraction vector $g$ = 100, which yield an uncertain Burgers vector of [001], [101], or [10$\bar{1}$].

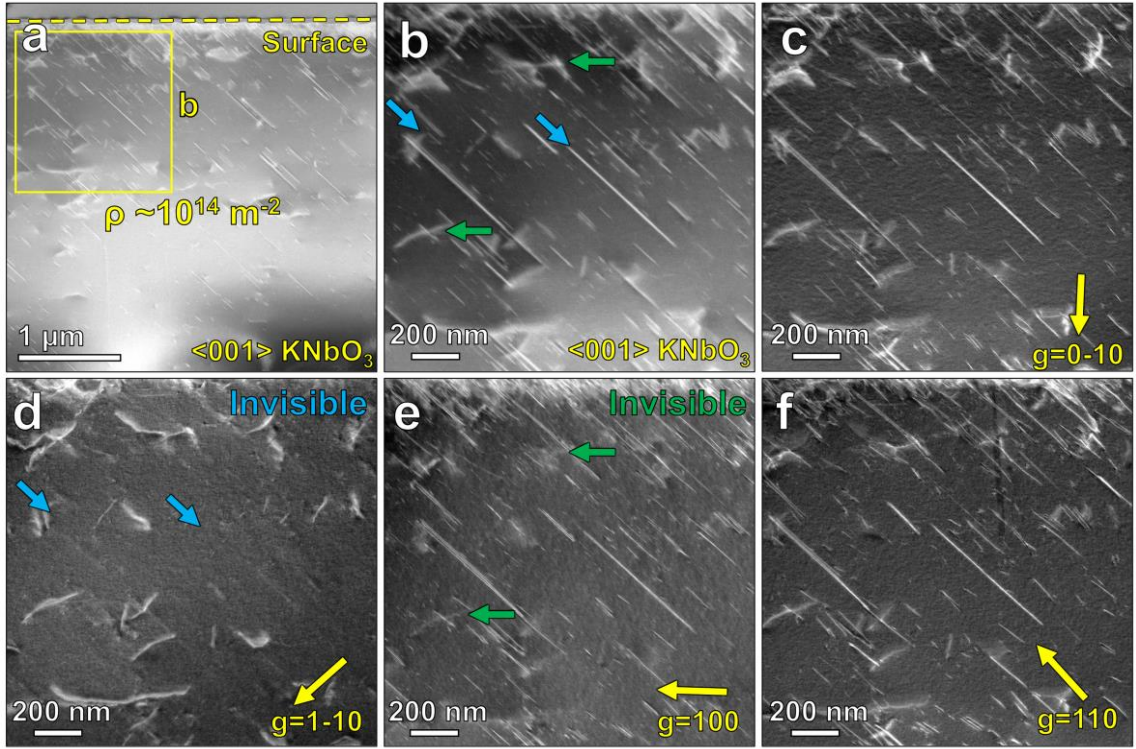

**Figure 1.** ADF-STEM images of dislocations in (001) KNbO$_3$ with 10× scratching under different diffraction conditions. (a) Dislocation distribution of the 10× scratched (001) KNbO$_3$ with a density of ~10$^{14}$ m$^{-2}$; change of the dislocation contrast along (b) <001> zone axis, (c) $g$ = 0$\bar{1}$0, (d) $g$ = 1$\bar{1}$0, (e) $g$ = 100, (f) $g$ = 110.

We then used the HAADF-STEM imaging, coupled with fast Fourier transform (FFT) and inverse FFT (IFFT), to investigate the core structures of the head or tail of the 45° dislocation segments, as illustrated in **Figure 2**. Here again, we use pseudo-cubic (pc) cell to index diffraction patterns in the orthorhombic (ortho) cell for clarity, namely, [110]$_{ortho}$ = [100]$_{pc}$, [1-10]$_{ortho}$ = [010]$_{pc}$, and [001]$_{ortho}$ = [001]$_{pc}$. At the head of the 45° dislocation dipole marked by the blue box in **Figure 2a**, we identified two edge-type dislocations with opposite Burgers vectors of 1/2a[110]$_{pc}$ and 1/2a[$\bar{1}\bar{1}$0]$_{pc}$, respectively (**Figure 2b**). The IFFT pattern in **Figure 2c** reveals the dislocation dipole was separated by a distance of multiple unit cells (~7 nm). At the tail of the single 45° dislocation segment marked by the green box in **Figure 2a**, we identified two climb dissociated edge dislocation with the same Burgers vector of 1/2a[$\bar{1}\bar{1}$0]$_{pc}$. The corresponding IFFT pattern in **Figure 2e** reveals that the climb distance is ~5 nm. Geometric phase analysis (GPA) map in **Figure 2f** generated by the FFT pattern insert demonstrates the strain field distribution around the two edge dislocations. Two selected $g$ vectors, $g$ = 100 and $g$ = 0$\bar{2}$0, marked with blue circles in the FFT pattern, were used to compute the strain field in the $x$ direction ($\varepsilon_{xx}$). The resulting strain map in **Figure 2f** revealed the areas of tensile strain (red) and compressive strain (blue) around the two edge dislocations, which exhibit an extremely high gradient.

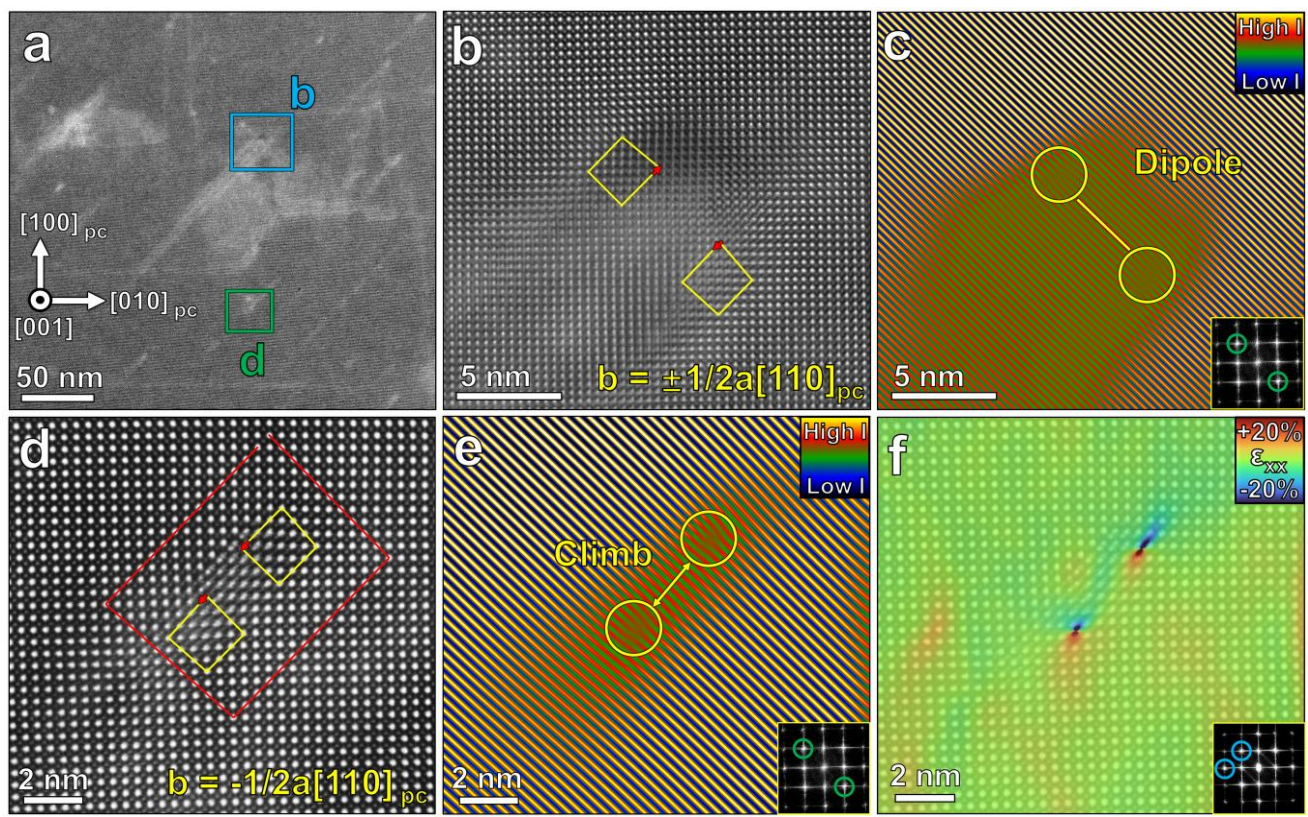

**Figure 2.** Dislocations in the 10× scratched (001) $KNbO_3$: (a) Low magnification of the 45°-inclined dislocation lines; (b) dislocation core structure of the head of the 45° dislocation dipole marked in the blue box in (a), which includes two edge dislocation with opposite Burgers vectors; (c) An IFFT pattern in (b) shows the atom planes; (d) dislocation core structure of the tail of the 45° dislocation line, containing two climb dissociated edge dislocation with same Burgers vector; (e) An IFFT pattern in (d) shows the atom planes; (f) A GPA result reveals the strain distribution around the dislocation cores in (d).

**Figure 3a** displays the $H_2$ yield using reference and dislocation-rich KNO crystals as the piezocatalysts. For a blank control group, the average $H_2$ yield is 0.84 ppm, demonstrating the negligible $H_2$ production performance of the reactor. The average $H_2$ yield of the dislocation-rich sample is 77.04 ppm, which is 2.46 times higher than that of reference sample. As shown in **Figure 3b**, after deducting $H_2$ mass produced by the reactor, the average $H_2$ production rate of reference and deformed sample is 1.84 and 4.61 μmol/h, respectively. **Figure 3c** presents the comparison on $H_2$ production rate between KNO single crystals with/without dislocations and other reported piezocatalysts. Due to a well-aligned domain structure[20], deformed KNO exhibits a higher $H_2$ production rate than other polycrystalline bulk ceramics. Compared to $BaTiO_3$ single crystal, a superior $H_2$ production performance of deformed sample can be attributed to its stability in polarization at high temperature. Since the Curie temperature $T_c$ of $BaTiO_3$ (~ 130 °C)[11] is lower than that of KNO (> 400 °C)[21], higher temperature caused by collapsed cavitation bubbles during ultrasonic process may lead to a depolarization in $BaTiO_3$ and slower $H_2$ evolution. Furthermore, it is promising that dislocation engineering can further lift the $H_2$ production rate to a higher level, even comparable to some powder-based catalysts like nanowires with high length-diameter ratio and piezoelectric potential[22].

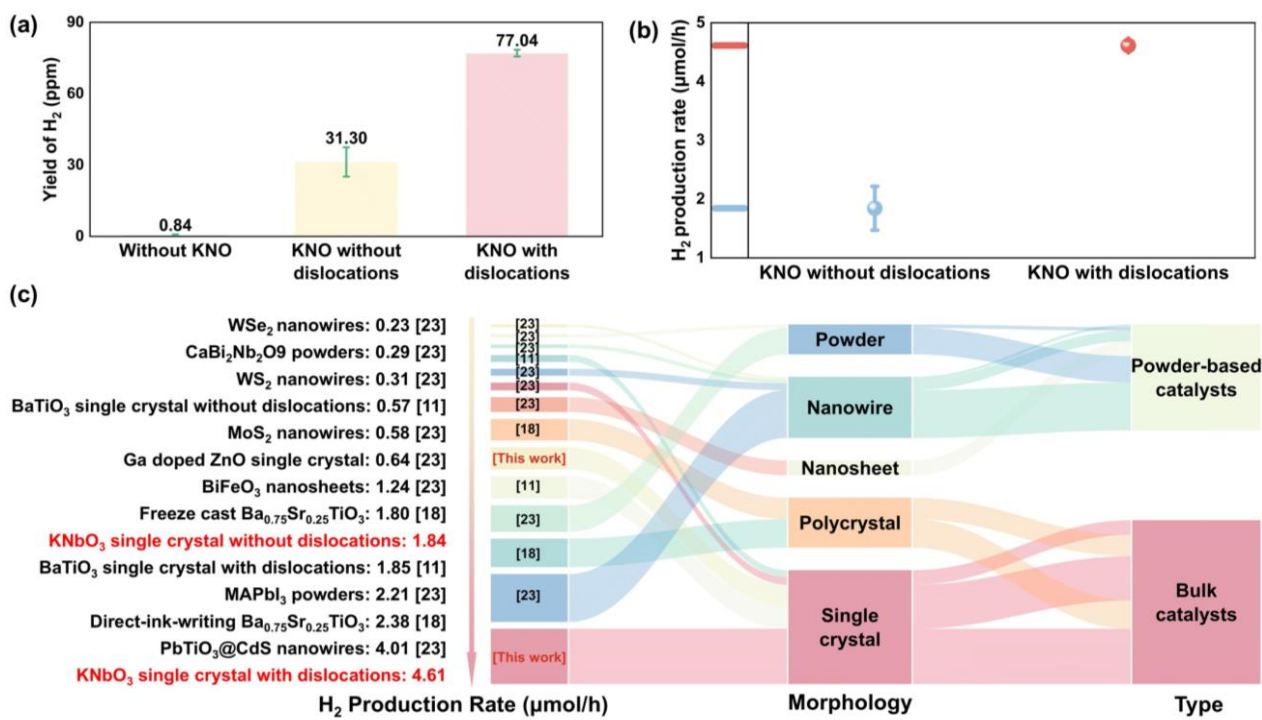

**Figure 3**. Performance and mechanism of piezocatalytic hydrogen production. (a) Yield of hydrogen, (b) hydrogen production rate in $KNbO_3$ single crystal, (c) a comprehensive comparison of piezocatalysts in hydrogen production rate[11,18,23].

We discuss in **Figure 4a** the possible mechanism based on screening the charge effect in piezocatalysis[24]. Initially, KNO remains electrically neutral as the bound charges induced by spontaneous polarization $P_1$ are compensated by oppositely charged external screening charges located at the two polar surfaces. When subjected to compressive stress, such as that induced by cavitation bubbles during ultrasonic treatment, the original charge equilibrium is disturbed. This leads to a lower polarization $P_2$ and the subsequent release of excess screening charges from the surface. These released charges will participate in redox reactions with surrounding species, including water molecules, dissolved oxygen, hydroxide ions ($OH^-$), and protons ($H^+$), until a new electrostatic equilibrium is established[18,24]. Upon removal of the external stress, the polarization is restored from $P_2$ to $P_1$, resulting in the increase of polarization-associated bound charges, followed by absorbing free charges from ambient environment and triggering new redox reactions. Subsequently, KNO crystal returns to the initial state. It should be noted that the difference between reference sample and deformed sample is the polarization. Previous study[25] confirms local polarization enhancement near a single dislocation. Furthermore, the higher spontaneous polarization $P_s$ has been obtained in (K, Na)NbO$_3$ single crystals with internal stress, consequently contributing to a larger local piezoelectric coefficient $d_{33}$[26]. Thus, compared to the reference sample (**Figure 4b**), the deformed sample (**Figure 4c**) exhibits a higher local strain $\varepsilon$ (see GPA results in **Figure 2f**). The high strain and stress fields in the vicinity of dislocations[27], as illustrated by the red region in **Figure 4c**, require additional bound polarization charges to compensate for the enhanced built-in electric field induced by the dislocations. This effect is particularly pronounced in the dislocation core region. Furthermore, previous work[11] has shown that dislocations can enhance electrical conductivity, resulting in faster release and accumulation of surface charges. As the schematic diagram shown in **Figures 4b-c**, KNO with dislocations can produce more hydrogen and oxygen, which benefits from its higher local polarization.

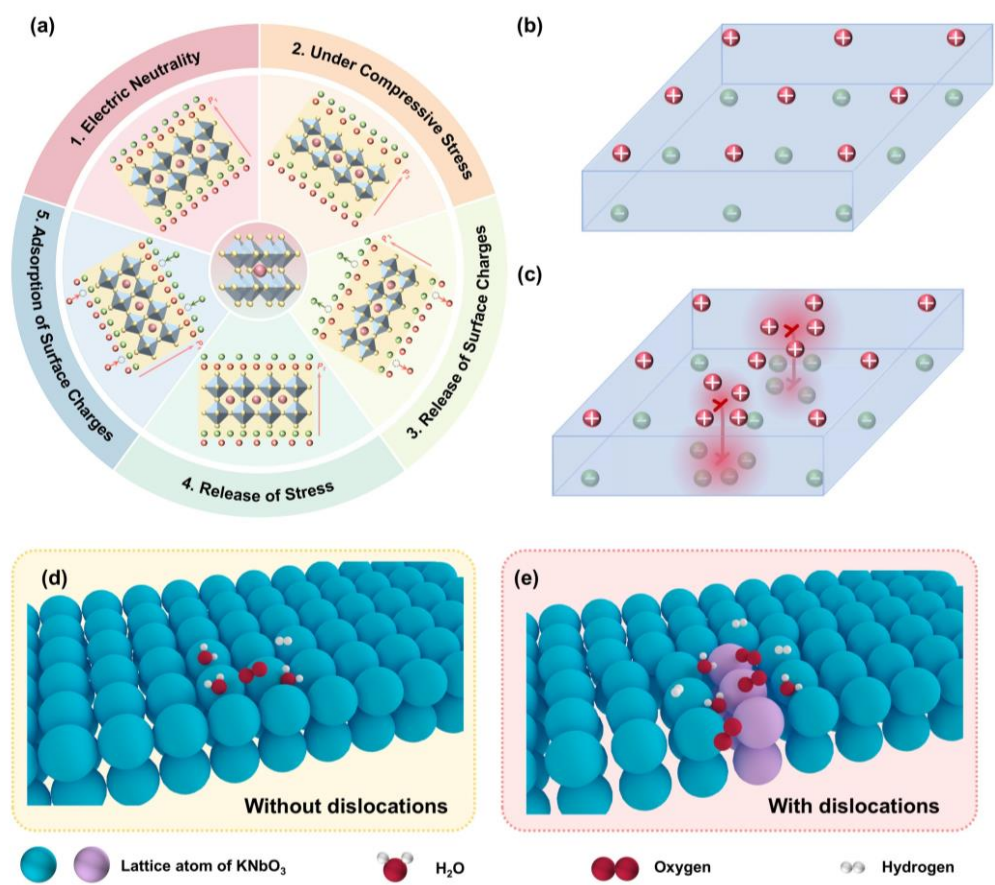

**Figure 4**. Mechanism of piezocatalytic hydrogen production. (a) Schematic of the screening charge effect. Schematic illustrations of KNbO$_3$ single crystals (b, d) without and (c, e) with dislocations, showing differences in surface charge distribution and water splitting process.

In summary, dislocations were successfully introduced into the near-surface region of bulk potassium niobate (KNbO$_3$) single crystals through room-temperature cyclic scratching, enabling a notable enhancement by larger than 2 times in piezocatalytic hydrogen production rate. Compared to conventional high-temperature imprinting, this method is more energy-efficient and avoids the risk of crack formation or structural degradation. The improved catalytic performance is proposed to be primarily attributed to the elevated polarization level induced by strain fields surrounding the dislocations, which promotes more efficient charge separation and accelerate surface redox reactions. These findings demonstrate that mechanical deformation-based dislocation engineering at room temperature provides a feasible and potentially scalable approach to enhance the catalytic functionality of piezoelectric materials without additional chemical dopants or compromising structural integrity.

**This file contains the following supplementary files:**

**Figure. S1**. Optical images of the large plastic zone (yellow rectangle) on the single-crystal $KNbO_3$ after Brinell indenter scratching at room temperature.

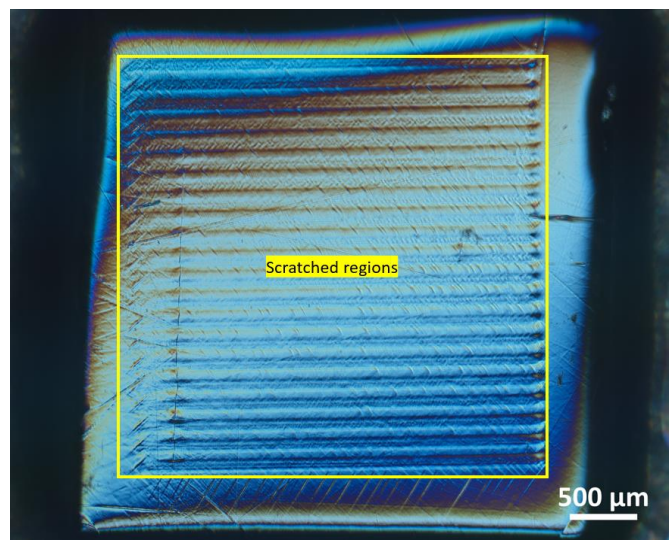

**Table S1.** Dislocation Burgers vector analysis in **Figure 2** in the main text.

|   |      | Burgers vector *b* | | | | | | | | |
|---|------|-----|-----|------|-----|-----|------|------|-----|-----|
|   |      | 010 | 100 | 1-10 | 110 | 101 | 10-1 | 01-1 | 001 | 011 |
| *g* | 0-10 | O | X | O | O | X | X | O | X | O |
|   | 100  | X | O | O | O | O | O | X | X | X |
|   | 110  | O | O | X | O | O | O | O | X | O |
|   | 1-10 | O | O | O | X | O | O | O | X | O |

*Note: Dislocations are invisible when **g·b**=0. X represents invisible, O represents visible*

**Acknowledgement**

X. Fang thanks the funding by European Research Council (ERC Starting Grant, Project MECERDIS, No.101076167). W. Lu acknowledges the support by Shenzhen Science and Technology Program (grant no. JCYJ20230807093416034), the Open Fund of the Microscopy Science and Technology-Songshan Lake Science City (grant no. 202401204), National Natural Science Foundation of China (grant no. 52371110) and Guangdong Basic and Applied Basic Research Foundation (grant no. 2023A1515011510). The authors also acknowledge the use of the facilities at the Southern University of Science and Technology Core Research Facility and State Key Laboratory of Powder Metallurgy, Central South University. The ECCI images by Dr. E. Bruder and helpful discussion and input by Prof. J. Rödel at TU Darmstadt are acknowledged.

**Author contributions**

X. Fang & Y. Zhang: conceptualization and project design. H. Gong, Y. Zhao, J. Zhang, Shan Xiang, Xiang Zhou, O. Preuß: experiments, data collection and analysis. W. Lu, X. Fang, Y. Zhang: supervision. H. Gong, X. Fang: writing, first draft. All: review and editing of the draft.

**Data availability statement**

The data used for this publication has been included in the manuscript or supplementary materials.

**Conflict of interest:** The authors declare no known conflict of interests.